\documentclass[jkps,preprint,fleqn,showpacs,showkeys]{revtex4}
\usepackage[pdftex]{graphicx}
\usepackage{amssymb}
\usepackage{amsmath}
\usepackage{bm}
\begin{document}
\title{Sleepless in Seoul: `The Ant and the Metrohopper'}
\author{Keumsook \surname{Lee}}
\email{kslee@sungshin.ac.kr}
\affiliation{Department of Geography, Sungshin Women's University, Seoul 136-742}

\author{Jong Soo \surname{Park}}
\affiliation{School of Computer Science and Engineering, Sungshin Women's University, Seoul 136-742}

\author{Hannah \surname{Choi}}
\affiliation{School of Engineering Sciences and Applied Mathematics, Northwestern University, Evanston, IL 60208, USA}

\author{M. Y. \surname{Choi}}
\affiliation{Department of Physics and Astronomy, Seoul National University, Seoul 151-747}

\author{Woo-Sung \surname{Jung}}
\email{wsjung@postech.edu}
\affiliation{Department of Physics and Basic Science Research Institute, Pohang University of Science and Technology, Pohang 790-784}

\date{\today}

\begin{abstract}
One of Aesop's (La Fontain's) famous fables `The Ant and the Grasshopper' is widely known to give a moral lesson through comparison between the hard working ant and the party-loving grasshopper.  Here we show a slightly different version of this fable, namely, ``The Ant and the Metrohopper," which describes human mobility patterns in modern urban life. Numerous real transportation networks and the trajectory data have been studied in order to understand mobility patterns. We study trajectories of commuters on the public transportation of Metropolitan Seoul, Korea.  Smart cards (Integrated Circuit Cards; ICCs) are used in the public transportation system, which allow collection of transit transaction data, including departure and arrival stations and time. This empirical analysis provides human mobility patterns, which impact traffic forecasting and transportation optimization, as well as urban planning.
\end{abstract}



\maketitle

Human trajectories are usually analyzed through the use of real world databases \cite{1,2,3,kslee,4,5,6,7,8,9,10,11} such as transportation networks, travel diary and bank note dispersion, as well as mobile phone movement, for use in traffic forecasting and transportation optimization \cite{12}, as well as urban planning \cite{13}. The Metropolitan Seoul Subway network operates a smart card system which keeps track of travel information on every passenger. Here, we analyze the passenger flows on a single day, based on the transaction data of the smart card on 24 June 2005. The transit transaction database, which holds data on over 10,000,000 transaction per day, contains time/position information on each passenger's travel. Therefore, it is possible to track the movements of individuals, because each smart card has its own ID. The transaction data cover 2,746,517 passengers who take the subway on the analyzed day. We find that modern urban life can be characterized well by the commute pattern on the public transportation system.

\begin{figure}
\includegraphics[width=1.0\textwidth]{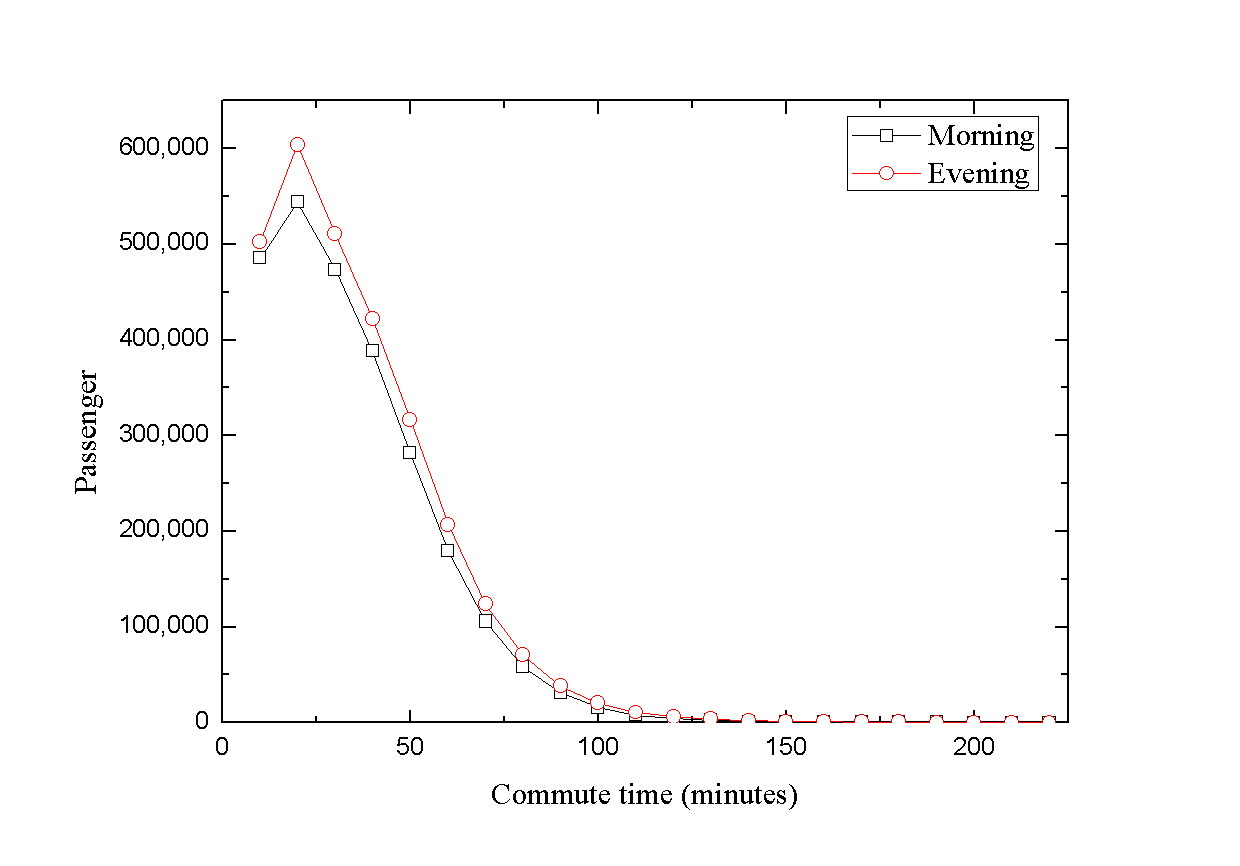}
\caption{(Color online) Commute time distribution in the morning and evening.
}\label{commutetime}
\end{figure}

\begin{figure}
\includegraphics[width=1.0\textwidth]{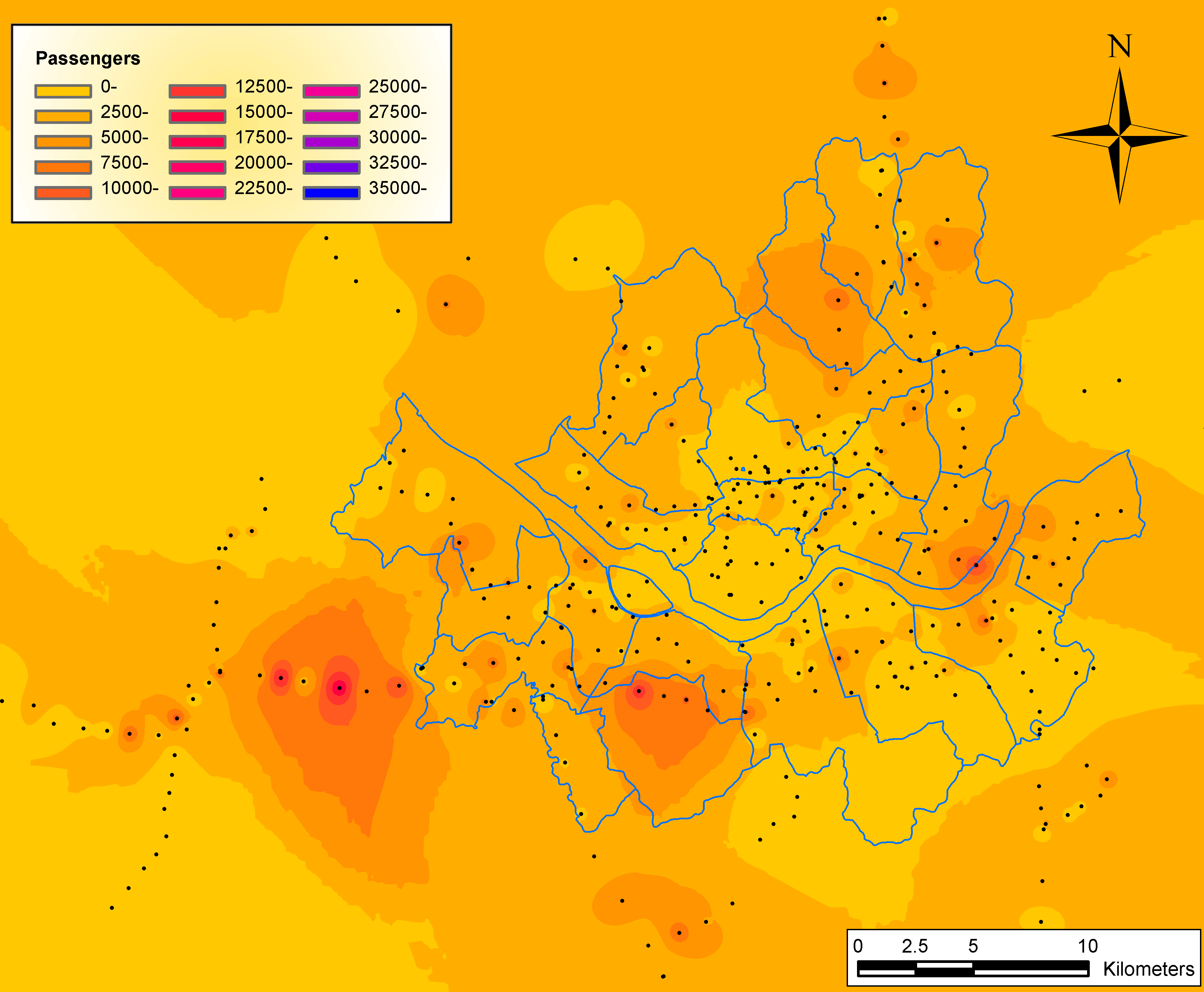}
\caption{(Color online) Spatial distribution of the departures in the morning. Black dots correspond to subway stations, and blue lines are ward boundaries in Seoul.
}\label{fig1}
\end{figure}

\begin{figure}
\includegraphics[width=1.0\textwidth]{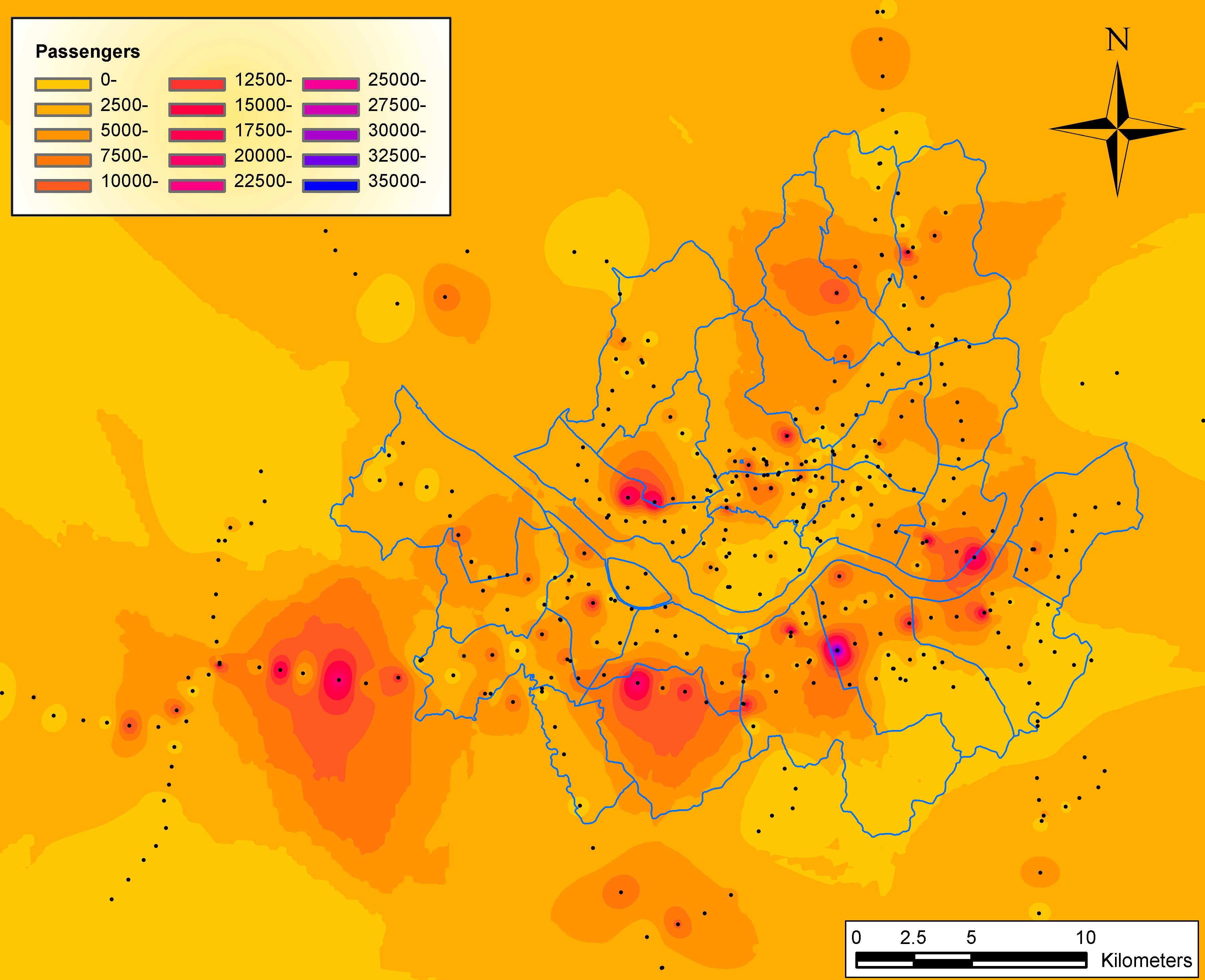}
\caption{(Color online) Spatial distribution of the arrivals in the evening.
}\label{fig2}
\end{figure}

The Metropolitan Seoul Subway system, consisting of nearly 400 stations, carries approximately three million passengers a day.  Among them, we analyze those trajectories that have more than two transaction data in one smart card in order to investigate the commute pattern. In this study, `morning' corresponds to `prior to 11 AM,' while `daytime' represents `between 11 AM and 5 PM' and `evening is `after 5 PM'. Figure \ref{commutetime} represents the commute time distribution in the morning and evening, while Figure \ref{fig1} exhibits the spatial distribution of the departures in the morning, obtained by considering with geographic information systems. A great portion of people live in the suburbs, so they must return home after work.  Surprisingly, many of them stop by somewhere else on their way home, so the spatial distribution of arrivals in the evening (Fig. \ref{fig2}) shows features different from those of the departure distribution in the morning.  In the evening, a large number of people visit such popular entertainment areas as Gangnam, Shinchon, Hongdae and Hyehwa, which display dense arrival distributions in Fig. \ref{fig2}.

We classify the commute patterns into three categories, according to the commutate among three places: home, office, and other locales.  Category 1 portrays `the Ant,' who has the endless cycle of getting up, going to work and returning home while Category 2 represents `the Metrohopper,' an urban version of `the Grasshopper,' who stops by other locales after work and does not take the subway to return home. The purpose of visiting the other locales by a Metrohopper can be either entertainment (as reflected by the areas in Fig. \ref{fig2}) or a secondary job. It is important to note that the Metropolitan Seoul Subway system operates daily from 5 AM to 1 AM.  It is likely that most of the Metrohoppers return home very late, by taxi or on foot, after the service hour of the subway. Category 3 corresponds to `the Hybrid,' a cross between `the Ant' and `the Metrohopper,' who stops in other locales after work but eventually returns home by subway. While 45\% of the passengers behave as ants, the metrohoppers and the hybrids, 55\% of the total, who are not confined only to work and home, characterize Seoul as a ``sleepless" metropolis (Table 1).

\begin{figure}
\includegraphics[width=1.0\textwidth]{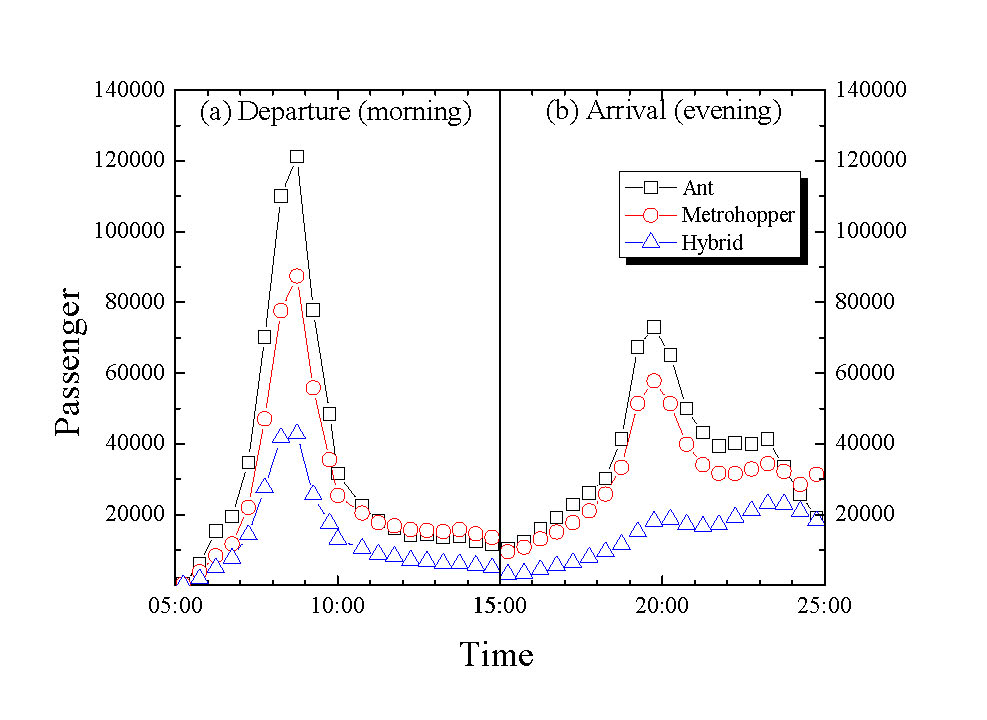}
\caption{(Color online) (a) Departure time distribution of the first trip and (b) arrival time distribution of the last trip of the day.
}\label{fig3}
\end{figure}

We also investigated the departure- and the arrival-time distributions.  Figure \ref{fig3} displays the distributions of (a) the departure times of the first trips and (b) the arrival times of the last trips for three categories: the Ant, the Metrohopper, and the Hybrid.  All three categories show similar departure-time distributions, which indicates that they leave home at similar times in the morning.  However, the arrival-time distributions for the Metrohoppers and for the Hybrids show two peaks while that for the Ants has only one peak with a shoulder.  The second peak in the former corresponds to the after-work activities.  Interestingly, the second peak for the Hybrids, who returning home by subway, is more conspicuous than that for the Metrohoppers, who take transportation other than subway.

Unlike the grasshopper in rural life in Aesop's fable, the metrohopper may not be a loser in urban life. Important business decisions and social gatherings, as well as self-development, tkae place at after-work sessions.  Particularly, after-work sessions in Korea often continue until midnight, as a sequel to day work.  We, thus, conclude that the celebrated fable `The Ant and the Grasshopper' has a different outlook in modern urban life.  Which of the two, the ant or the metrohopper, will be more effective in urban life?  This is left as an open question.



\newpage
\clearpage

\newpage
\clearpage

\newpage
\clearpage

\newpage
\clearpage

\newpage
\clearpage
\begin{table}
\begin{tabular}{c|cc}
Category&Passengers&Percent of total\\
\hline
1 (Ant)&771,935&45.03\%\\
2 (Metrohopper)&653,046&38.09\%\\
3 (Hybrid)&289,349&16.88\%\\
\hline
Total&2,746,517&100.0\%\\
\end{tabular}
\caption{Numbers of passengers in each category.}
\label{table1}
\end{table}

\end{document}